\documentclass[12pt,preprint]{aastex}
\usepackage{graphicx,natbib}

\newcommand{\etal}{et al.\ }

\shorttitle{Stellar masers} 
\shortauthors{Kemball}

\begin{document}

\title{Stellar masers, circumstellar envelopes, and supernova remnants}

\author{Athol J. Kemball}
\affil{University of Illinois at Urbana-Champaign, USA}

\begin{abstract}
This paper reviews recent advances in the study or circumstellar
masers and masers found toward supernova remnants. The review is
organized by science focus area, including the astrophysics of
extended stellar atmospheres, stellar mass-loss processes and
outflows, late-type evolved stellar evolution, stellar maser
excitation and chemistry, and the use of stellar masers as
independent distance estimators. Masers toward supernova remnants are
covered separately. Recent advances and open future questions in this
field are explored.
\end{abstract}

\keywords{masers, stars:AGB and post-AGB, stars:atmospheres, supergiants,
planetary nebulae, supernova remnants}

\section{Introduction}

This review concerns circumstellar masers and the masers associated
with supernova remnants. The material covered focuses primarily on
results published since the last maser symposium, IAU 206
\citep{Mig02}, but avoids those areas covered separately by other
papers in these proceedings.

In considering circumstellar masers, it is appropriate to start by
first examining the host stars and the basic properties of the most
common masers found in their circumstellar shells. The late-type
evolved stars that host circumstellar masers are either red
supergiants or large-amplitude long-period variables (LALPV), such as
thermally-pulsing asymptotic giant branch (TP-AGB) stars. These stars
are astrophysically important for well-known, but still important,
reasons \citep{Hab96}:

\begin{enumerate}

\item{Through their mass-loss they enrich the chemical and dust
composition of the interstellar medium from which new stars form.}

\item{They are high-luminosity tracers of faint and obscured stellar
populations and their main sequence progenitors; intermediate mass
stars, for example, achieve their highest brightness at the tip of the
AGB.}

\item{They highlight an important short-lived stage of stellar
evolution in the transition from the AGB to planetary nebulae.}

\end{enumerate}

The key properties of the most common circumstellar maser transitions
are summarized in Table~\ref{tab:masers}. Numerical values in this
table are approximate only; representative temperatures and
densities are drawn from \citet{Rei02}. In the most basic model of
circumstellar masers, the maser species enumerated in
Table~\ref{tab:masers} are assumed concentrically distributed about
the central star, at radii broadly correlated with the transition
excitation temperature.

\begin{table}\def~{\hphantom{0}}
  \begin{center}
  \caption{Common masers in late-type, evolved stars}
  \label{tab:masers}
  \begin{tabular}{lccc}\hline
                     & SiO   &   H$_2$O & OH \\\hline
  Transition         & ~${v=0,1,2,3,..; \triangle J=1}$~ & $6_{16} - 5_{23}$ & ~~~${}^2\Pi_{3/2}, J=3/2, \triangle F=\{0,1\}~~~$\\
  Radius (AU)        & $\sim$ 3-6   & $\sim 100$ & $\sim 1000$\\
  $n_H$ (cm$^{-3}$)  & $5.10^{10}$   & $10^8$ & $10^7$\\
  Temperature (K)    & 1500 & 750 & 450\\
  Pumping mechanism  & Radiative or & Collisional & Radiative\\
                     & collisional  &             &          \\\hline
  \end{tabular}
 \end{center}
\end{table}

Given the late-type evolved stars and the basic properties of
circumstellar masers they host, it is appropriate to ask the broader
question of what larger astrophysical questions can be explored by
stellar maser studies.  We can answer this in part by noting first
that masers are unique probes of obscured circumstellar astrophysics
due to both their high brightness temperature and compact spatial
structure. As such, they give us unique insights into the extended
stellar atmosphere, the region between the photosphere and dust
formation point. They also allow the study of the mass-loss process
both near the star and in the outflow at larger radii. The masers 
allow an exploration of circumstellar excitation and pumping
conditions as well as circumstellar chemistry. They also allow
independent methods of stellar astrometry and distance measurement. We
consider each of these scientific areas in more detail below.

In addition to the topics listed here, stellar masers provide insights
into stellar magnetic fields, basic maser theory, surveys for galactic
population and dynamical studies, and are also used in specialized
galactic center dynamics studies. Though each is extremely important,
these issues are not included further in this review as they are
covered by separate papers in these proceedings. Masers toward
supernova remnants are covered in Section 7 below.

\section{Extended atmosphere}

\begin{figure}[b]
\begin{center}
\includegraphics[height=3.072in,width=4in,angle=0]{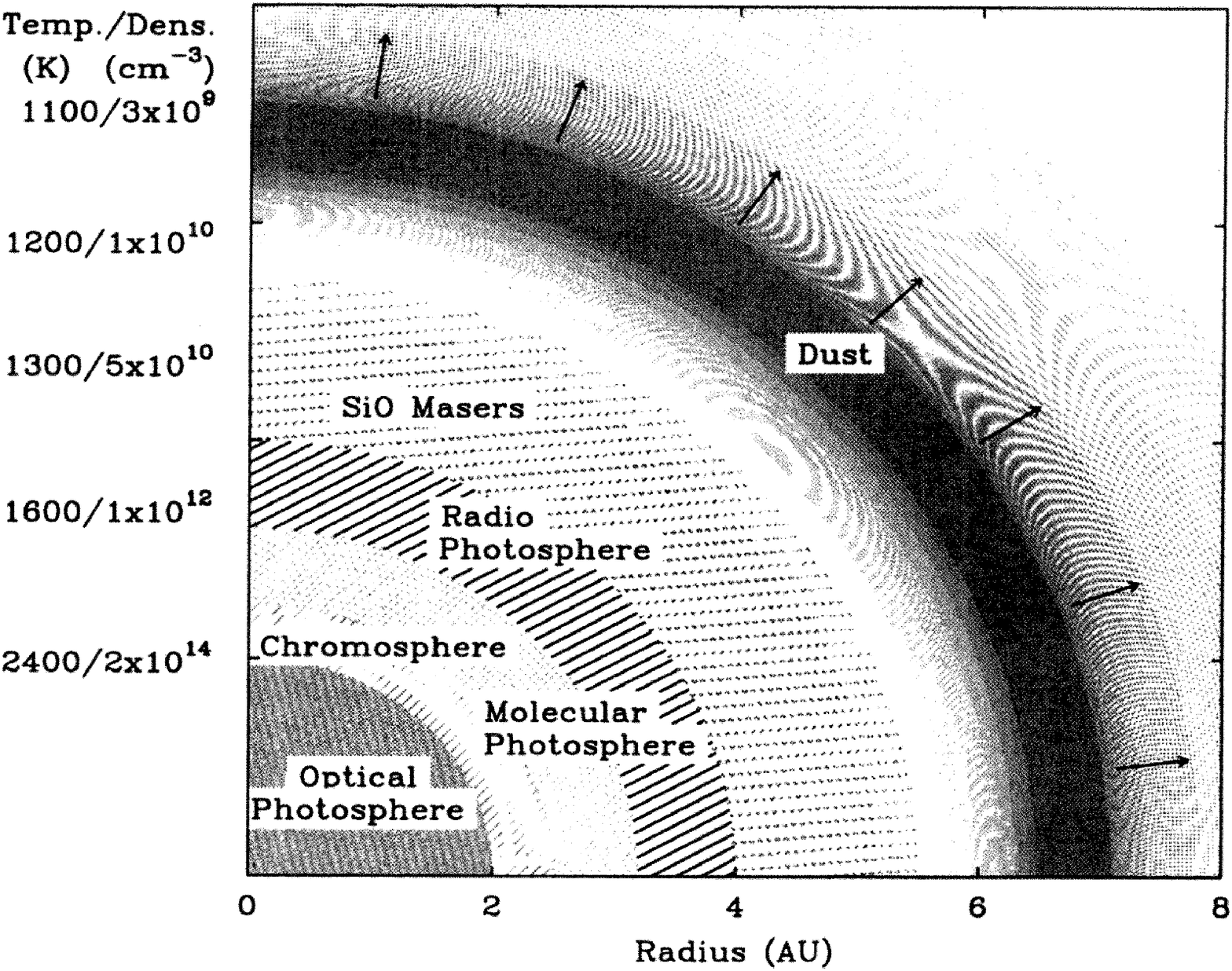}
  \caption{Schematic representation of the circumstellar envelope of a
  stellar maser from the inner envelope out to the dust formation
  zone; Figure 3 from \citet{Rei02}}
   \label{fig:reid02}
\end{center}
\end{figure}

The near-circumstellar environment is shown schematically in
Figure~\ref{fig:reid02}, taken from \citet{Rei02}. This region is
dominated by mass-loss, shocks from the centrally-pulsating star, and
is expected to have a complex kinematic and dynamical structure
\citep{Eli92}. High-resolution imaging of this region is essentially
only provided by SiO maser observations; a capability unmatched by
contemporary telescopes in other wave-bands.

A number of astrophysical sub-themes are of interest and importance in
studies of the extended atmosphere. These can be broadly categorized
as issues of kinematics, dynamics, and the physical conditions
supporting individual maser transitions. In kinematics, there are open
questions concerning localized flows, their connection to the central
star and convective envelope, the nature and extent of any global
asymmetry established at the base of the mass-loss process, and the
matching of gas kinematics in this region to pulsation hydrodynamic
models. The dynamical issues concern the nature of any shaping forces,
and the magnitude and dynamical influence of local or global stellar
magnetic fields. Physical conditions cover issues of excitation and
chemistry.

Knowledge of the convective stellar envelope is limited primarily to
numerical simulation studies \citep{Fre02,Por97}. In the 3-D
heat convection study of a red-giant atmosphere by \citet{Por97}, in
which heat was supplied steadily to the central core to drive the
convection, two unexpected behaviors arose, namely pulsation and
large-scale convective flows. Complex convective structures and
phenomena in the stellar envelope are likely to have a significant
impact on the extended atmosphere.

The extended atmosphere can be studied uniquely using VLBI
observations of SiO masers, which act as ultra-compact astrophysical
probes of this region. Their importance as probes of this inner CSE
was realized soon after their first detection in the 1970s but early
imaging studies faced difficult technical challenges
\citep{Mor79,Gen79}, primarily due to the low sensitivity of
millimeter-wavelength VLBI arrays. The advent of improved VLBI
sensitivity enabled new observations \citep{Col92,Miy94}; in particular,
the commissioning of the VLBA\footnote{http://www.vlba.nrao.edu}, a
little over ten years ago, provided dramatically-improved imaging
capabilities compared to the challenges faces by the earliest
observations \citep{Dia94,Gre95}.

A particularly important result in the last several years has been the
conduct of simultaneous near-infrared (NIR) interferometry and radio
interferometry imaging campaigns of stellar SiO maser sources using
the VLBA both in conjunction with the
VLTI\footnote{http://www.eso.org/projects/vlti} \citep{Bob05,Wit05} and with
IOTA/FLUOR\footnote{http://tdc-www.harvard.edu/IOTA}
\citep{Cot04,Cot06}. These observations have allowed the
unambiguous co-location of photospheric and SiO radii at
${R_{SiO}\over{R_*}} \sim 1.5-3.0$, without having to rely on
model-based photospheric diameter estimates. The monitoring campaign
reported by Cotton et al. (2004) also hints at a possible relationship
between the relative location of the 3.6 $\mu m$ photosphere and a
dependence of the SiO maser radius on the relative 3.6 to 2.2 $\mu m$
photospheric radii. Those stars in this monitoring sample with higher
3.6 $\mu m$ opacity in the molecular photosphere also appear to have
large dust condensation radii.

At longer IR wavelengths, the
ISI\footnote{http://isi.ssl.berkeley.edu/index.htm} has recently
reported an important direct measurement of radial pulsation in the
diameter of Mira at 11 $\mu m$, showing a fluctuation in the continuum
photosphere of +11\% between stellar pulsation phase $\phi=-0.08$ to
0.15 \citep{Wei03}, in good agreement with theoretical models for
Mira-class pulsation.

Pulsation affects the extended atmosphere directly, driving periodic
shocks into the near-circumstellar environment (NCSE)
\citep{Ber85}. The shock emerges at pre-maximum, accelerating gas
outwards, which subsequently falls back under the influence of gravity
\citep{Hin82,Hin97,Alv00}. This produces a double-lined S-shaped
velocity profile in photospheric infrared lines such as 1.6 $\mu m$ CO
$\triangle \nu=3$ absorption \citep{Hin82}. The shocks levitate
material above the the hydrostatic atmosphere and subsequent radiation
pressure on dust couples to the gas and carries it further outwards.
Shock formation in the NCSE is studied in a number of
spherically-symmetric, piston-driven hydrodynamic models of the
stellar atmosphere \citep{bow88,Bes96,Hum02}. Shock propagation
through the NCSE in these models leads to a saw-tooth velocity profile
with radius. The models also accurately predict a net mass-loss of
material to the outer circumstellar environment over time and a
complex interaction between accelerating and decelerating gas from
successive pulsation cycles.

In the time since the last maser conference in 2001, the number of
synoptic imaging campaigns of stellar SiO masers has increased
significantly and yielded new results. These imaging campaigns are
enabled by the technical advances in VLBI imaging at this wavelength,
as noted above. Cotton and collaborators are monitoring a small LALPV
sample in a snapshot imaging campaign \citep{Cot04,Cot06}. Synoptic
imaging of the Mira variable TX Cam has also continued both at
short-spacing in pulsation phase \citep{Dia03,Gon05}, and over longer
phase increments \citep{Yi05}. \citet{Gon05} has shown that the median
SiO maser component lifetime in TX Cam is approximately 150-200 days,
a substantial fraction of the 557 day stellar pulsation period. This
permits detailed proper motion studies, as reported by
\citet{Dia03,Gon05}. A global proper motion analysis shows outer
components falling back from earlier pulsation cycles simultaneously
with outflow during the current pulsation cycle. This is consistent
with the saw-tooth velocity profile discussed above; however, there
are significant local departures from a globally ordered flow in the
measured proper motions in NSEW quadrants of the projected shell
\citep{Dia03}. The mean velocities are broadly consistent with shock
damping in the radio photosphere however, as deduced from upper limits
to radio continuum stellar variability by \citet{Rei97}.

An important astrophysical question in this area is the degree of
agreement between measured mean global radial motions and the
predictions of pulsation hydrodynamic models of the type discussed
above. Recent reported mean global motions are summarized in
Table~\ref{tab:prop} against the pulsation phase interval $\triangle
\phi$, and whether the outflow was generally contracting, expanding,
or neither. This table show that there are disagreements between
individual stars in the same pulsation phase range (e.g. R Aqr, TX
Cam, and VX Sgr), where both expansion and contraction are variously
reported. However, note that this is comparing a known binary, a Mira
variable, and a supergiant. Some authors \citep{Cot04,Cot06} have also
reported no systematic motions with pulsation phase for their
sample. However, it is clear that this picture is likely more
complex. If we examine TX Cam over multiple pulsation periods we can
argue for inter-cycle variability and possible interaction between two
competing time-scales, the pulsation period and the free-fall
gravitational time-scale, possibly causing a non-strictly repeating
pattern of global motions. We also note that there are systematic
measurement effects that need to be carefully addressed. The inner
shell is not circular and robust estimators are needed to determine
the mean inner radius \citep{Dia03}. For at least one pulsation phase
interval of TX Cam, there is clear evidence for ballistic deceleration
with physically reasonable parameter values \citep{Dia03}. It is not
true for all pulsation phase cycles however.

\begin{table}\def~{\hphantom{0}}
  \begin{center}
  \caption{Measured pulsation kinematics}
  \label{tab:prop}
  \begin{tabular}{llccl}\hline
  Reference     & Sources  &   $\triangle \phi$ &  Projected  & Outflow or\\
                &          &                    &  velocity (kms$^{-1}$) & infall \\\hline
  Boboltz \etal\ (1997)    & R Aqr    & 0.78-1.04          &  -4.2 & Contraction\\
  \citet{Dia03} & TX Cam   & 0.7-1.5            &  +7   & Expansion\\
  \citet{Cot04} & See below$^1$  & Various &    & No pattern\\
  \citet{Yi05}  & TX Cam   & 0.60-1.05          &     & Slowing expansion\\
  \citet{Gon05} & TX Cam   & 1.5-2.7            &     & Both \\
  \citet{Cot06} & See below$^2$ & Various  &    & No pattern\\
  \citet{Che06} & VX Sgr   & 0.75-0.80          &  -4   & Contraction\\\hline
\multicolumn{5}{l}{1: R And, Mira, U Ori, R Leo, W Hya, S Crb, U Her, R Aqr, R Cas.}\\
\multicolumn{5}{l}{2: Mira, U Ori, R Aqr.}\\
  \end{tabular}
 \end{center}
\end{table}

The issue of rotation in the extended atmosphere is also important
from a kinematic and evolutionary perspective. Two detections of SiO
maser shell rotation, including differential rotation, have been
reported since the last maser meeting, namely for R Aqr \citep{Hol01}
and Mira \citep{Cot06}, both known binaries. This is, however, not
common for other sources, as evident in the sector-averaged velocity
plots reported by \citet{Cot06} for their LALPV sample. Substantial
rotation is difficult to explain in evolved late-type stars that are
not part of multiple star systems.

\begin{figure}[b]
\begin{center}
\includegraphics[height=3.16in,width=3in,angle=0]{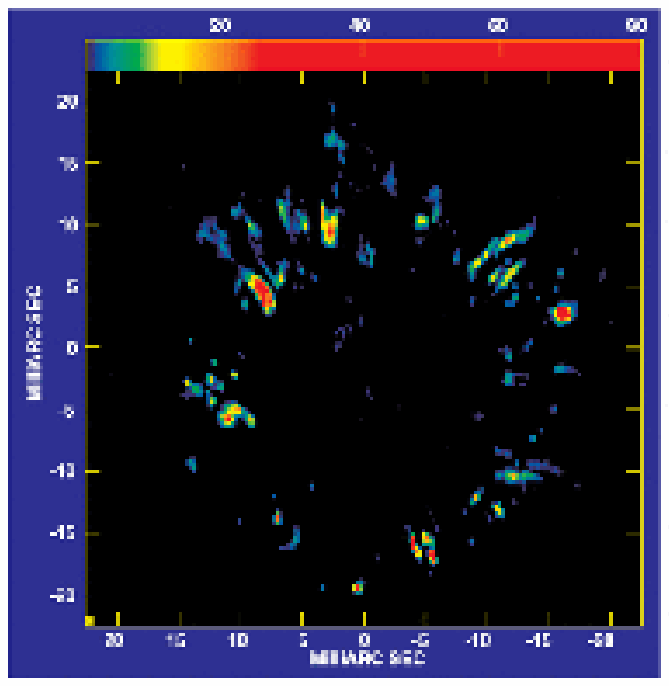}
  \caption{VLBI maps of $v=1, J=1-0$ SiO maser emission toward TX Cam,
  summed over all channels of emission. The bar shows the flux
  scale in Jy/beam (Figure 3 from \citet{Yi05}).}
   \label{fig:yi2005}
\end{center}
\end{figure}

The expanded range of SiO imaging studies, such as those enumerated in
Table~\ref{tab:prop}, continue to reveal clear evidence for
spatially-coherent arcs and filaments in the SiO maser region
\citep{Kem97}. These local features have recently been argued by
\citet{Yi05} to be from spokes of SiO flows with line-of-sight
velocities either slightly red- or blue-shifted with respect to the
systemic stellar velocity. This kinematic structure maximizes the
line-of-sight velocity coherence. The radial spokes are especially
visible in several TX Cam image epochs reported by \citet{Yi05}, one
of which is reproduced here as Figure~\ref{fig:yi2005}.

\section{Outflows}

The water and hydroxyl masers probe intermediate and outer
circumstellar radii \citep{Cha86,Boo81}. These masers trace the
mass-loss process beyond the dust formation point; in this region, the
flow is driven by radiation pressure on dust. The astrophysical
sub-themes that are relevant to this region of the outflow include the
kinematic structure of the mass-loss, including the degree of
asymmetry on local and global scales, and the connection of this
kinematic pattern to the inner NCSE. For cases where a dominant
kinematic pattern is clear, the science issue becomes that of
identifying any shaping processes. This is also a particularly
important area of the outflow in which to explore the evolutionary
onset of more extreme asymmetry related to post-AGB evolution,
including stellar jets.

Some of these science questions have been explored in recent
MERLIN\footnote{http://www.merlin.ac.uk} observations of stellar water
masers; \citet{Bai03} report the first connected-element resolution of
individual maser spots, and derive important parameters relating to
the water maser region surrounding four LALPV stars: IK Tau, U Ori, RT
Vir, and U Her. They report significant departures from spherical
symmetry in tori that have an inner radius of 6-16 AU and an outer
radius of 24-54 AU. Individual maser components are resolved out at
2-4 AU, and are found to be density-bounded $(\sim 10^{1-2}$ times
ambient), with a volume filling factor $< 0.5\%$. In related MERLIN
observations of the red-supergiant, VX Sgr, \citet{Mur03} have
performed a proper motion study of the water maser emission toward
this source over a five year interval. A bi-conical outflow is used to
describe the water maser kinematics. A similar over-density of maser
components and a comparable extrapolated cloud size at the stellar
surface of $\sim 0.07 R_*$ is reported, relative to that given by
\citet{Bai03} for the sample of low-mass AGB stars. Both observations
confirm an accelerating flow across the water maser region, with a
logarithmic velocity gradient of $0.5 \leq \epsilon \leq 1$.

Concerning the role of water maser studies as a probe of early-onset
post-AGB asymmetry, a particularly important result in recent years has
been the discovery and exploration of a rare sub-class of AGB stars
showing highly-collimated stellar jets \citep{Ima04,Vle06}. The
dynamical age of the jets is very young ($< 35-100$ years) and only a
handful of such sources are known, consistent with this dynamical
age. This rare class of ``water fountain'' sources is described in
separate review papers in these proceedings by Imai and Vlemmings.

A particularly unusual and rare source studied by \citet{Bab06,Szc06}
presents water masers in a warped disk around a companion to the SiC
star V778 Cyg.

OH observations of the supergiant NML Cyg are also reported by
\citet{Eto04}. In this source, they present evidence for two shells,
an older spherical shell, presumed from an earlier mass-loss phase,
and a younger asymmetric flow.

\section{Late-type stellar evolution}

In this section, we consider the issue of post-AGB evolution, and how
stellar masers can inform our understanding of this phase of stellar
evolution. Planetary nebulae (PNe) show a rich diversity of point- and
axi-symmetric structure \citep{Bal02}. The immediate post-AGB (PAGB)
evolution is heavily shrouded and masers provide a powerful method to
study this transition phase, complementary to imaging studies in other
wave-bands, such as infrared.

The astrophysical sub-themes that are important in this area concern
how to explain the prevalence of point- and axi-symmetry in PNe. Four
major shaping mechanisms have been proposed including binarity,
equatorial density enhancement, a globally-acting magnetic field and
sculpting by highly-collimated stellar jets \citep{Mor03}. It is of
significant interest to determine when the shaping mechanism is
activated on the AGB. 

An over-arching issue in this area is also how to identify and select
post-AGB objects in samples of OH-IR stars, such as the complete 1612
MHz OH survey conducted by \citet{Sev97a,Sev97b,Sev01}. PAGB
identification in such surveys is considered by \citet{Sev02} based on
clustering in color-color IR plots, using both IRAS and MSX
data. Follow-up observations of candidate PAGB stars in the Sevenster
survey are presented by \citet{Dea04}; their observations show a
significantly higher incidence of asphericity in the candidate PAGB
sample and further strengthen the IR color-color selection analysis by
\citet{Sev02}.

Kinematic models for OH outflow sources are presented by \citet{Zij01}
using a two component model, consisting of a shell/torus (20-25
kms$^{-1}$) and higher-velocity linear outflow (10-80 kms$^{-1}$).

Masers are particularly rare in young PNe, and only a handful of such
sources are known although they are very important. These sources are
discussed by G\'omez (these proceedings).

\section{Excitation and chemistry} 

The question of maser excitation, pumping, and circumstellar chemistry
cuts across both theory and observation. It is needed to
understand physical conditions in the masing regions and most
importantly to properly interpret observations. In addition, these
issues are central to predicting new masing transitions that may be
accessible using future telescopes.

Both collisional \citep{Loc91} and radiative \citep{Deg76} mechanisms
have been proposed for SiO maser pumping. There are observational
discriminants between the two pumping mechanisms, primarily the
relative spatial location of v=1 and v=2 masers, variability studies,
and linear polarization morphology. There have been important recent
observational results in this area, but conclusions drawn in the
literature remain mixed. Based on the location of the v=1 and v=2
J=1-0 masers toward TX Cam, \citet{Yi05} cite evidence for collisional
pumping. However, \citet{Sor04} draw the opposite conclusion from v=1
and v=2 J=1-0 imaging of the SiO masers toward IRC+10011 and $\chi$
Cyg. In recent variability studies, \citet{Par04} reports finding a
zero phase lag between IR and SiO flux density over an 11-year short
time-spacing monitoring campaign, and \citet{McI06} finds no time-lag
between several vibrationally-excited J=1-0 transitions toward
Mira. Both authors argue for radiative SiO pumping as a result. 
\citet{Ase05} consider the effect of anisotropic radiative pumping
and the Hanle effect on the linear polarization morphology of SiO
stellar masers.

Studies of the relative location of different rotational transitions
in the same vibrational state, primarily J=2-1 relative to J=1-0, have
shown perplexing results \citep{Sor04,Phi03}. In the handful of
sources imaged so far, the morphology is significantly different,
contrary to basic theoretical expectations. Line overlaps may strongly
influence the theoretical predictions about spatial location however,
as discussed by \citet{Sor04}.

Another important recent result in SiO maser excitation has been the
imaging of v=0, J=1-0 SiO maser emission towards a sample of late-type,
evolved stars by \citet{Bob04}. A predominance of weak masers is
confirmed and one source was found to have thermal
emission. Although their spatial resolution is limited, they propose
that v=0 is found at twice the radius of the v=1 emission.

Several new possible masers have been reported toward C-stars since
our last meeting, including new HCN lines toward a sample of carbon stars
\citep{Bie01}, and a possible weak OH maser detection toward
IRC+10216 \citep{For03}.

\citet{Men06} also report the possible new detection of maser action in the
H$_2$O $v_2=1,6_{61}-7_{52}$ line at 294 GHz toward VY CMa. In
addition, \citet{Tho03} report weak maser emission in the J=9 $l$-type
HCN transition toward CRL 618.

\section{Independent distance estimates}

Stellar masers can also be used to provide independent stellar
distances. This technique, demonstrated by \citet{vLa00}, is
based on the assumption that the compact, blue-shifted maser emission
is assumed coincident with the central star. The technique was
applied to further AGB stars by \citet{Vle03}. 

OH stellar astrometry is likely limited to distances less than 1 kpc
\citep{Vle03}; astrometry using H$_2$O and SiO masers can reach larger
distances. Recently, \citet{Kur05} demonstrated stellar water maser
astrometry in observations of UX Cyg. \citet{Vle02} also showed positional
coincidence between the brightest water maser feature toward U Her and
the stellar Hipparcos position.

\section{1720 MHz OH masers toward supernova remnants}

OH 1720 MHz masers, shock-excited toward supernova remnants (SNR), were
first detected by \citet{Gos68}. Their phenomenology is now well-known
and a sound theoretical basis has been established for their modeling
and pumping \citep{Gre02}.

They are an important probe of SNR shock interactions with molecular
clouds, and are also found in the galactic center circumnuclear
disk. They are considered in more detail in separate invited papers in
these proceedings by Brogan, Yusef-Zadeh, and Hewitt.

\section{The future}

There remain important open astrophysical problems in the area of
circumstellar and SNR masers, requiring advances in both observation
and theory. An integrated model for the kinematics and dynamics of the
mass-loss process from the convective stellar envelope, through the
extended atmosphere to the outer circumstellar shell will require
further detailed observations, and more complex numerical hydrodynamic
models in order to explain the observed local and global asymmetries
and to identify the underlying shaping forces. This is particularly
important in so far as it concerns stellar magnetic fields of
late-type, evolved stars. Recent detections of stellar jets in water
maser emission also open new windows on post-AGB evolution. The theory
of astrophysical masers continues to remain central to this field, and recent
observations have posed new theoretical questions, particularly in the
area of SiO excitation, that are vital to proper interpretation of
observations. We are also on the cusp of major new observational
advances in this field, driven by the commissioning of
VERA\footnote{http://veraserver.mtk.nao.ac.jp},
ALMA\footnote{http://www.alma.nrao.edu}, and the
eVLA\footnote{http://www.aoc.nrao.edu/evla} and look forward to
significant future results from these instruments.

\begin{acknowledgments}
The author would like to gratefully acknowledge the support of the
National Science Foundation under grant NSF AST 05-07473.
\end{acknowledgments}

\end{document}